\documentclass[twocolumn,showpacs,preprintnumbers,amsmath,amssymb]{revtex4}

\usepackage{graphicx}
\usepackage{dcolumn}
\usepackage{bm}

\begin{document}


\title{Thermal lensing-induced bifocusing of spatial solitons in Kerr-type optical media}
\author{Alain~M.~Dikand\'e}
\affiliation{Laboratory of Research on Advanced Materials and Nonlinear
Sciences (LaRAMaNS), Department of Physics, Faculty of Science, University of
Buea PO Box 63 Buea, Cameroon}
\date{\today}
\begin{abstract}
Thermo-optical effects cause a bifocusing of incoming beams in optical media,
due to the birefringence created by a thermal lens that can resolve the incoming beams into two-component signals of different
polarizations. We propose a non-perturbative theoretical description of the process of formation of double-pulse solitons in Kerr optical media with a 
thermally-induced birefringence, based on solving simultaneously the heat equation and the propagation
equation
for a beam in a one-dimensional medium with uniform heat flux load. By means
of a non-isospectral Inverse Scattering Transform assuming an initial solution
with a pulse shape, a one-soliton solution to the wave equation is obtained
that represents a double-pulse beam which
characteristic properties depend strongly on the profile of heat spatial distribution.
\end{abstract}

\maketitle
\begin{figure} 
\includegraphics[width=3.0in]{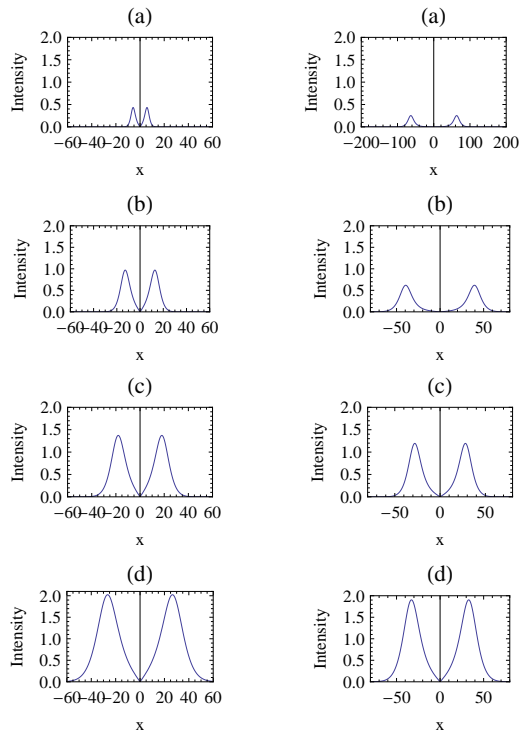}
\caption{\label{fig:one} $x$-axis profiles of $\vert q(x,z) \vert$ for $z=0.5$
(left graphs) and $z=10$ (right graphs). Values of $\alpha$ are (a) $0.1$, (b)
$0.05$, (c) $0.01$, (d) $0.005$}
\end{figure}
\begin{figure} 
\includegraphics[width=3.0in]{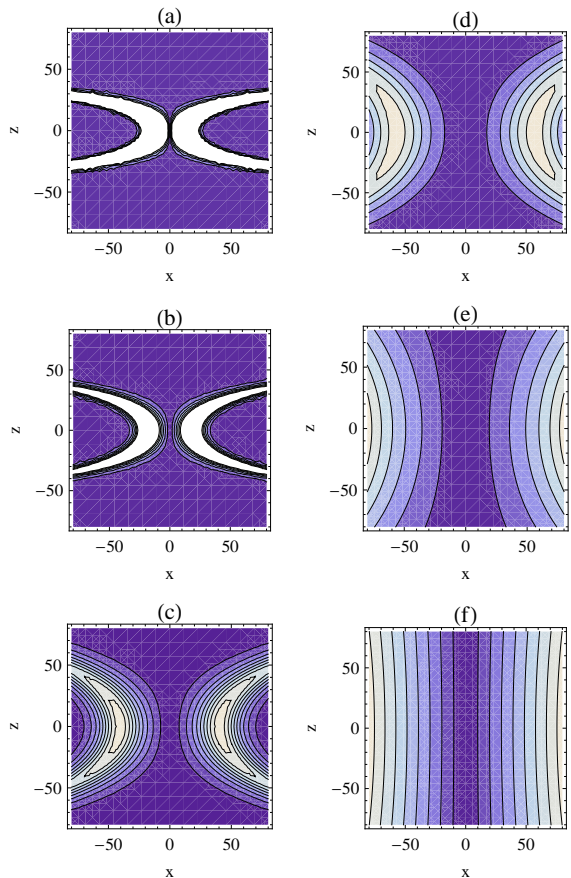}
\caption{\label{fig:two} Contour plots of $\vert q(x,z) \vert$ for different
$\alpha$ listed as: (a) $0.01$, (b) $0.005$, (c)$0.001$, (d) $0.0005$, (e)
$0.0001$, (f) $0.00001$.}

\end{figure}

In dielectric media with thermo-optical
effects, the modulations of incoming beams can lead to a wavefront
distortion~\cite{dist1,dist3,dist4}
reflecting their instability. Generally
this instability gives rise to a depolarization of a high-power
field~\cite{hans,menzel1,moshe1,moshe2,wang,resolv2}
due to a thermally-induced birefringence which is
attributed~\cite{biref1,biref2,biref3} to a
change in the refractive index of the medium. It has been
established~\cite{dist1} that
this change in the refractive index originates from heat deposition in the
propagation medium, resulting in a space-dependent
temperature gradient (so-called thermal lensing)~\cite{dist1,hans,koechner}. For
media with linear indices,
the thermal lens leads to a drastic change in the
irradiance along 
the beam axis so that the resulting depolarization can strongly degrade 
the beam quality requiring thermal lensing compensation. However in nonlinear 
media such as Kerr media, the nonlinearity can be a relevant
self-compensaton factor~\cite{alex} stabilizing 
multi-wave modes 
generated by the thermal birefringence.
\\
While several materials exhibiting thermo-optical effects are known in the
literature~\cite{dist1}, a most investigated one is the
solid-state laser Nd:YAG~\cite{dist1,hans,koechner,eich}. This material is
represented as a rod crystal with a cylindrical geometry, where the change in
temperature induces thermal distortion of
incoming laser beams~\cite{resolv2}. 
For this particular material, several theoretical attempts have been made to
formulate the spatial profile of the temperature gradient 
along the rod exploiting available experimental data. In particular, in
refs.~\cite{dist1,koechner,eich} it was found that in 
the cylindrical rod configuration where the heat is
generated at a constant rate~\cite{koechner,resolv2}, a 
quadratic spatial distribution gives a very good
description of the experimentally
observed birefringence and the resulting beam
bifocusing~\cite{resolv2,maik}.  \\
But thermo-optical processes are actually common to a broad class of
materials,
not just solid-state lasers. Indeed, 
photonic crystals and optical fibers (including laser fibers) displaying
nonlocal
thermal and photothermal properties have been 
considered in the recent past, from both
experimental and theoretical points of view~\cite{chen,martha}. These
materials share in 
common the fact that heat resulting from the input pump source
causes physical variation of the material. Namely, the material expands
with the heat load due to a stress gradient formed which produces
space-dependent birefringence in the material. Quite remarkably when
the thermal gradient is strong enough the thermally-induced
birefringence can resolve a polarized high-power input beam into two-component
beams~\cite{hans,resolv2,resolv1}, moreover upon recombination after
traversing the bulk the 
two beams with two polarization directions are no longer in phase with one
another such that the polarization
state of the
input beam cannot be recovered~\cite{dist4,resolv1}. \\
The thermally-induced depolarization phenomenon together with the resulting
beam bifocusing have been investigated experimentally and theoretically 
by several authors~\cite{hans,menzel1,moshe1,moshe2,wang,resolv2}. In photonic
crystals with a nonlinear Kerr-type
dielectric susceptibility 
competing with diffraction, nonlocal thermal properties of the propagation
medium have very recently been shown to induce twin-mode spatial solitons 
so-called dipole solitons~\cite{yak,cohen,skupa,bambi}. The robustness of these
double-pulse structures provides strong indication that 
the competition between nonlinearity and thermal lensing might be a stabilizing
factor for the double-polarized modes, potentially
observable in several optical materials including solid-state and optical fibers
with thermally-induced birefringence.\\
The present work aims at proposing a non-perturbative description
of the generation of a double-polarized pulse beam in optical materials with
thermally-induced linear
birefringence, paying particular attention to nonlinear optical media 
with a quadratic spatial profile of temperature distribution. Our objective is
to provide a consistent theoretical framework for understanding better the
effects of
thermal lensing on beam propagation in nonlinear optical materials, of which the
YAG fiber
laser which temperature profile was exprimentally
established~\cite{dist1,koechner,eich}.
We will first demonstrate, by solving the heat equation, that the topography
of temperature distribution is quadratic in the case of a one-dimensional ($1D$)
anisotropic material uniformly
loaded
at a constant heat flux rate. Next, using the relationship between the temperature gradient 
and the induced inhomogeneous optical index, 
we shall derive an effective refractive index for the 
material by combining the thermally-induced optical index and
a Kerr-type 
refractive index reflecting the intrinsic nonlinearity of optical properties of
the 
$1D$ anisotropic material. With the help of this effective refractive index we shall formulate
the 
propagation equation for beams in the thermal nonlinear medium. Note that this equation is a 
Nonlinear Schr\"odinger equation (NLSE) with a repulsive
external
quadratic potential, that has already been derived in ref.~\cite{cari} but solved
perturbatively. Here instead, a 
non-perturbative treatment will be proposed based on a non-isospectral Inverse
Scattering Transform (IST) method~\cite{rada,dika1} with emphasis on
IST's initial solution being single-pulse shaped. \\
Thermal lensing can be described in simple words as a thermo-optical process
associated with a weak absorption of an
input beam that
induces a nonzero temperature gradient across the material, leading to a spatial
variation of its refractive
index~\cite{dist1,koechner,cohen}. Recent experimental as well as theoretical
developments on this process suggest that the
underlying mechanism involves a local refractive index change $\Delta n(x)$
which increases linearly~\cite{cohen,eich} with the temperature change $\Delta
T(x)$ i.e. $\Delta n= \beta \Delta T$, where $ \beta=dn/dT$ refers to the
thermo-optic coefficient~\cite{cohen,sowade}.
Thus
when the optical
beam of a uniform thermal load gets slightly absorbed and heats the material,
this produces heat that is conveyed by the electromagnetic wave. If $\rho$
denotes the uniform heat flux density and $\kappa$ the
heat conductivity coefficient, the heat diffusion in the material along a
preferred direction
(for $1D$ anisotropic materials of current interest), driven by the
uniform heat flux load, is determined
by the heat equation:
\begin{equation}
\kappa\,\nabla^2 T(x)= -\rho. \label{a1}
\end{equation}
Since eq.~(\ref{a1}) is key to the current analysis we consider its most
general solution given by:
\begin{equation}
T(x)= -\frac{1}{2}a_2\,x^2 + a_1\,x + T_0, \label{a1a}
\end{equation}
where $a_2=\rho/\kappa$, $a_1$ and $T_0$ being two arbitrary real constants.
Formula~(\ref{a1a}) is
consistent with the quadratic law of temperature variation found for most laser
fibers in the presence of a uniform 
head load. More specifically in YAG fiber lasers this law
is common~\cite{dist1,koechner,eich} and is consistent with the optical
bifocalility associated with a thermally-induced 
birefringence, that promotes double-polarized laser beams from an input laser
field. \\ 
For the sake of simplicity we require
the temperature gradient to be zero
and temperature to take a bare value $T_0$ at $x=0$ (i.e. the ambient
temperature). The change of temperature
in the material along the $x$ axis, hereafter assumed to be the axis of beam
propagation, then reads:
\begin{eqnarray}
\Delta T(x)&=& T_0 - T(x) \nonumber \\  
          &=& \frac{1}{2}a_2\,x^2. \label{a1b}
\end{eqnarray}
With the last formula we derive the following expression for the local
refractive index change:
\begin{equation}
\Delta n(x) =\frac{1}{2}\alpha\,x^2, \hskip 0.2truecm \alpha
=\rho\,\beta/\kappa. \label{a2}
\end{equation}
Now if the intrinsic optical properties of the material are dominated by
Kerr-type phenomena, the homogeneous part of the refractive index
can be expressed as $(k^2c/2\pi)n_2\, I$ where $I$ is the beam intensity. With
the help
of~(\ref{a2}) we can readily define an effective refractive index $n(x)$ for the
thermal nonlinear material viz:
\begin{equation}
n[x,\,I]= \Delta n(x) + (k^2c/2\pi) n_2\, I. \label{a3}
\end{equation}
Assuming that the wave motion is fast along the axis of anisotropy (i.e. $x$)
but very slow along $z$~\cite{cohen}, the 
paraxial approximation on the $2D$ wave equation for an electromagnetic field
$q(x,z)$ leads to: 
\begin{equation}
\frac{\partial^2 q}{\partial x^2} + 2ik \frac{\partial q}{\partial z} +
n[x,\,I]\, q= 0. \label{a4}
\end{equation}
As already stressed eq.~(\ref{a4}) is actually not new, indeed the same equation
was obtained~\cite{cari} for the same problem but solved 
following the collective-coordinate method. To this last point, eq.~(\ref{a4}) is an inhomogeneous NLSE and so can in principle be solved using the
collective-coordinate method. However this is a perturbative
method and consequently requires that the thermal lensing
is sufficiently weak, so that the Kerr nonlinearity remains the main
governing factor in the modulation and stability of signals in the
thermal nonlinear medium. So to say any
input beam sent in the medium must be modulated into a signal of
permanent single-pulse shape, with eventually an acceleration or slow down of
the
pulse due to the 
thermal lensing. In fact this
consideration is very far from any acceptable consistency with the physics of
the
process under study, which the double-polarization of the incident beam is a
most salient aspect.     \\ 
Being interested in a solution to eq.~(\ref{a4}) which is more consistent with
experiments we try for a non-perturbative approach. Remark
to start that this equation can be
rewritten in the following form:
\begin{equation}
\frac{\partial^2 q}{\partial x^2} + 2ik \frac{\partial q}{\partial z} +
\left[(k^2c/2\pi) n_2\,\vert q\vert^2 - V(x)\right]q= 0, \label{a5}
\end{equation}
corresponding to a NLSE with an external potential
\begin{equation}
V(x)=-\alpha\,x^2/2 \label{pot}
\end{equation}
which is quadratic in $x$ with a maximum at $x=0$. The physics behind this
quadratic potential is contained both in its profile and the parameter $\alpha$
defined in~(\ref{a2}), which determines the strength of the
thermal birefringence on the beam shape. One remarkable side of this
physics emerges from the assumption of the heat flux density $\rho$ and the heat
conductivity $\kappa$ as being fixed, such that $\alpha$ appears to be 
increasing with an increase of the thermo-optic coefficient $\beta$. Thus when
$\beta$ is
increased the curvature of the quadratic profile of heat distribution in the
material is more and more pronounced so that 
the effect of thermal lensing on beam modulation is more and more strong. In
fact the external potential is expected to be much effective 
on the beam position and according to the profile of this potential given
by~(\ref{pot}),
the centre of mass of the input beam should experience a trapping
force from 
the manifestly  expulsive quadratic potential.\\
When $\alpha$ is large enough such that the potential field erected by the
thermal lensing process on the beam path is strongly localized, its contribution
must be fully taken into consideration. In this last respect we follow
the non-isospectral IST technique proper~\cite{rada,dika1} to equations of this
specific kind, 
considering an initial signal $q(x,z=0)$ of a permanent single-pulse shape
along $x$ at $z=0$.
For eq.~(\ref{a4}) this technique leads to the
following one-soliton solution:
\begin{equation}
q(x,z)=Q_0(x,z)\,sech\,[\Psi(x, z)]\,e^{i\Phi(x,z)}, \label{a6}
\end{equation}
with
\begin{equation}
Q_0(x,z)=\sqrt{\frac{2\pi}{cn_2}}\,\frac{x\,f(z)}{k}, \label{a6a}
\end{equation}
\begin{eqnarray}
\Psi(x, z)&=&\ln\left(2\vert f(z)\vert\right) - 2V(x)f(z) + \Psi_0,\label{a7a}
\\
\Phi(x, z)&=& 2V(x)g(z) + \Phi_0, \label{a7b}
\end{eqnarray}
\begin{equation}
f(z)= \frac{sech^2(z/\zeta)}{1 + 8(\rho\,\beta/\kappa)^2\,\tanh^2(z/\zeta)},
\label{a6b}
\end{equation}
\begin{equation}
g(z)= \frac{\kappa^2 +
8(\rho\,\beta)^2}{2\kappa\sqrt{2\rho\,\beta\,\kappa}}\,\frac{\tanh(z/\zeta)}{1 +
8(\rho\,\beta/\kappa)^2\,\tanh^2(z/\zeta)}, \label{a6c}
\end{equation}
\begin{equation}
\zeta=(2/\alpha)^{1/2}k. \label{a6d}
\end{equation}
If the "sech" function in~(\ref{a6}) reminds a pulse signal, the complicated
dependence of its 
prefactor $Q_0(x, z)$ in $x$ and $z$ clearly suggests an actually complex pulse
structure 
for the beam on the $xz$ plane. To gain insight about
what this dependence implies for the signal profile, in fig.~\ref{fig:one} we
plotted the beam
amplitude $\vert q(x,z) \vert$ as a
function of $x$ at two different positions in the direction $z$ transverse to
the beam propagation, and for four distinct values of $\alpha$.
More explicitely the four left graphs in fig.~\ref{fig:one} represent $\vert
q(x,z) \vert$ versus $x$ at
$z=0.5$ for $\alpha=0.1$, $0.05$, $0.01$ and $0.005$, while the left graphs
represent $\vert q(x,z) \vert$ versus $x$ 
for same values of $\alpha$ but at $z=10$. As one sees, the signal intensity is
a two-component pulse which intensities are strongly 
dependent on the strength of the thermal lens potential.
It is quite noticeable on exploring the six figures 
that when $\alpha$ increases, the intensities of the two-component pulse
increase while
their width at half tails diminish. The last behaviour is
consistent
with the $\alpha$ dependence of parameter $\zeta$ defined in~(\ref{a6d}), which
indeed represents the average spatial extension of the pulse along the $x$
axis.\\
Another relevant feature emerging from graphs of fig.~\ref{fig:one} is the fact
that when $\alpha$ is decreased for a fixed value of $z$, the two constituents
pulses in the double-polarized beam 
preserve their shapes but their peak positions (i.e. centres) are gradually
shifted. The last feature is more transparent in the left graphs
corresponding to a relatively large value of $z$. In fact, the last behaviour
can be interpreted in terms of a ring profile for the signal intensity
in the $xz$ plane as reflected by the contour plots of fig.~\ref{fig:two}, where
shadows of the double-pulse
soliton in the $xz$ plane are represented for different values of $\alpha$.

The figure clearly indicates an increase of the separation between pulses in the
double-pulse signal, 
implying that the radius and curvature of the ring signal in the $xz$
plane are fixed by the magnitude of $\alpha$. \\
The double-pulse structure obtained as well as its ring profile emerging in $2D$ are reminiscent of dipole and ring-vortex 
solitons, predicted recently in some nonlocal nonlinear
media~\cite{cohen,bambi,fei}. However in these previous studies the physical origins of nonlocalities
were generally not well specified, whereas in
our context the quadratic form of the optical inhomogeneity has an
experimental foundation~\cite{dist1,koechner,eich}. Also, a previous
attempt~\cite{cari} to model the same problem led to
eq.(\ref{a4}). In this previous work the equation was treated perturbatively and so 
results could not reflect the remarkable aspects underlying the physics of thermal lensing  in Kerr media. As end remark, the 
universality of the IST one-soliton
solution~(\ref{a6}) can be checked by applying
any other exact spectral method to the generating equation, such as the Darboux method~\cite{nakka} combined with a Lax-pair
formalism with non-conserved spectral parameters~\cite{porse,khawa1}. \\ \\
Part of this work was done at the Abdus Salam International Centre for Theoretical Physics (ICTP) Trieste, Italy. I thank Markus Muller, Matteo Marsili, M. Kiselev and V. Kravtsov for their kind hospitalities. \\
 

\begin{thebibliography}{}\label{sec:references}
\bibitem{dist1}W. Koechner, {\em Solid State Laser Engineering}
(Springer-Berlin, $6th$ edition, 2006).
\bibitem{dist3} J. D. Foster and L. M. Osterink, J. Appl. Phys. {\bf 41}, 3656 (1970).
\bibitem{dist4}C. A. Klein, Opt. Eng. {\bf 29}, 343 (1990).
\bibitem{hans}H. J. Eichler, A. Hasse, R. Menzel and A. Siemoneit, J. Phys. D {\bf 26},
1884 (1993).
\bibitem{menzel1}M. Ostermeyer, G. Klemz and R. Menzel, Proc. SPIE {\bf 4629}, 67
(2002).
\bibitem{moshe1}I. Moshe and S. Jackel, J. Opt. Soc. Am. B {\bf 22}, 1228 (2005).
\bibitem{moshe2}G. Machavariani, Y. Lumer, I. Moshe, A. Meir, S. Jackel and N.
Davidson, Appl. Opt. {\bf 46}, 3304 (2007).
\bibitem{wang}Y. Wang, K. Inoue, H. Kan and S. Wada, J. Phys. D {\bf 42}, 235108
(2009).
\bibitem{resolv2}Y. Wang, H. Kan, T. Ogawa and S. Wada, J. Opt. {\bf 13}, 015703
(2011).
\bibitem{biref1}E. A. Khazanov, O. V. Kulagin, S. Yoshida, D. B. Tanner and D.
H. Reitze, IEEE Quantum Electron. {\bf 35}, 1116 (1999).
\bibitem{biref2}F. Genereux, S. W. Leonard and H. M. Driel, Phys. Rev. B {\bf 63},
161101(R) (2001).
\bibitem{biref3}I. Shoji, Y. Sato, S. Kurimura, V. Lupei, T. Taira, A. Ikesue
and K. Yoshida, Opt. Lett. {\bf 27}, 234 (2002).
\bibitem{koechner}W. Koechner, Appl. Opt. {\bf 9}, 2548(1970).
\bibitem{alex}A. S. Koujelev and A. E. Dudelzak, Opt. Eng. {\bf 47}, 085003(2008).
\bibitem{eich}H. J. Eichler, A. Haase, R. Menzel and A. Siemoneit, J. Phys. D {\bf 26}, 1884 (1993).
\bibitem{maik}M. Frede, R. Wilhelm, M. Brendel, C. Fallnich, F. Seifert, B.
Willke and K. Danzmann, Optics Express {\bf 12}, 3581 (2004).
\bibitem{chen}S. Vasudevan, G. C. K. Chen and B. S. Ahluwalia, Opt. Lett. {\bf
33}, 2779 (2008).
\bibitem{martha}M. Andika, G. C. K. Chen and S. Vasudevan, J. Opt. Soc. Am. B
{\bf 27}, 796 (2010).
\bibitem{resolv1}S. Wielandy and A. L. Gaeta, Phys. Rev. Lett. {\bf 81}, 3359(1998).
\bibitem{yak}A. I. Yakimenko, Y. A. Zaliznyak and Y. Kivshar, Phys. Rev. E {\bf 71},
065603 (2005).
\bibitem{cohen}C. Rotschild, O. Cohen, O. Manela, M. Segev and T. Carmon, Phys.
Rev. Lett. {\bf 95}, 213904 (2005).
\bibitem{skupa}S. Skupin, M. Saffman and W. Kr\'olikowski, Phys. Rev. Lett. {\bf 98},
263902 (2007)
\bibitem{bambi}F. Ye, Y. V. Kartashov, Bambi Hu and L. Torner, Opt. Lett. {\bf
34}, 2658 (2009).
\bibitem{cari}A. Gharaati, P. Elahi and S. Cari, Act. Phys. Pol. A {\bf 112},
891 (2007) 
\bibitem{rada}R. Balakrishnan, Phys. Rev. B {\bf 32} 1144 (1985).
\bibitem{dika1}A. M. Dikand\'e, J. Math. Phys. {\bf 49}, 073520 (2008).
\bibitem{sowade}R. Sowade, I. Breunig, C. Tulea and K. Buse, Appl. Phys. B {\bf
99}, 63 (2010).
\bibitem{fei}F. Ye, B. A. Malomed, Y. He and Bambi Hu, Phys. Rev. A {\bf 81},
043816 (2010).
\bibitem{nakka}Z. Xu, L. Li, Z. Li, G. Zhou and K. Nakkeeran, Phys. Rev. E {\bf 68}, 046605( 2003).
\bibitem{porse}R. Rhada, V. R. Kumar and K. Porsezian, J. Phys. A {\bf 41}, 315209 (2008).
\bibitem{khawa1}U. Al Khawaya, J. Phys. A {\bf 39}, 9679 (2006) \& A {\bf 42}, 265206 (2009).
\end{thebibliography}
\end{document}